# An auto-locking ultra-stable laser with sub-hertz linewidth

D. Jiao[1,3,4], G. Xu[1,3,4], J. Gao[1,2,3], X. Deng[1,3], J. Liu[1,3], Q. Zang[1,2,3], X. Zhang[1,2,3], R. Dong[1,3*], T. Liu[1,3**], and S. Zhang[1,3]

[1]National Time Service Centre, Chinese Academy of Science, Xi'an 710600, China

[2] University of Chinese Academy of Science, Beijing, 100039, China

[3] Key Laboratory of Time and Frequency Standards, Chinese Academy of Science, Xi'an, 710600, China

[4]Both the authors contributed equally to this work

E-mail: taoliu@ntsc.ac.cn; dongruifang@ntsc.ac.cn;

**Abstract**

We report in detail the design process and performance of an auto-locking ultra-stable laser with sub-hertz linewidth at the first time. The laser frequency is automatically stabilized to an optical reference cavity with a home-made controller, which is based on a combination of digital circuit and analog circuit. The digital circuit is used for diagnosing and manipulating the state of the ultra-stable laser, and the analog circuit is used for demodulating the discriminate signal and servo control. A method of searching the transmission signal in the closed-loop state instead of the open-loop state is proposed to reduce the locking time and improve the reliable of the auto-locking ultra-stable laser. The median time of 16.6s is obtained after 157 times of relocking, and the probability of less than 20 s is more than 86%. The median linewidth of 1.08 Hz is obtained, and the fractional frequency instability is less than $3.4 \times 10^{-15}$ at integration time between 0.1 and 100 s. The performance of this system demonstrates that will be used as an important subsystem to transfer the optical clock signal.

Ultra-stable lasers with high frequency instability are in great demands for many fields like high-precision coherent phase transfer through fiber links,[1-5] gravity wave detection,[6-7] fundamental physics tests,[8-9] and optical atomic clock.[10-11] For these attractive applications, many groups have made lots of excellent advances in the development of ultra-stable lasers by locking a laser to a Fabry-Perot cavity with the Pound-Drever-Hall (PDH) technique.[12-16] Nevertheless, most of them have been constrained to operate in well controlled laboratory environments. There is growing requirement in frequency-stable laser capable of operation outside the laboratory for applications such as geodesy,[17-18] hydrology,[19] and low phase noise microwave synthesis.[20] Hence, it's significant to investigate ultra-stable lasers capable of operating automatically in unmanned environment for a long time. Meanwhile, to ensure the date continuity it requires the ultra-stable laser to be relocked as fast as possible after losing locking. The most important part of such an ultra-stable laser is a controller that is used for diagnosing and manipulating the operating state of the ultra-stable laser, and locking ultra-stable laser quickly. Many efforts have been made in designs of auto-locking ultra-stable lasers.[21-23] In 2016, an digital controller for laser frequency stabilization has been implemented on field programmable gate arrays (FPGA) and programmed with LabVIEW software, the actual re-locking process lasts for 8 minutes for an ultra-stable laser stabilized to a 5000-finesse-cavity.[22] The commercial auto-locking ultra-stable laser has been developed by Stable-Laser-System (SLS) company and Menlo-system company, the frequency instability are less than $7\times10^{-15}$ within 100 s averaging time and less than $3\times10^{-15}$ at 1 s averaging time respectively, yet there is no detailed report on the auto-locking performance and the principle scheme of the ultra-stable laser.[23-24]

In this paper, we report in details of a frequency auto-locking ultra-stable laser. The controller of this ultra-stable laser is developed with a combination of digital circuit and analog circuit. The digital circuit is used for monitoring operation state and adjusting parameter. The analog circuit is used for demodulating the discriminate signal and servo control. Usually, the work flow of locking ultra-stable laser is complicated, including turning on triangular wave, searching for transmission signal, turning off triangular wave, starting servo control and optimizing the parameters, which is not conducive to the fast locking of the ultra-stable laser. Thus, an approach of searching for transmission signal in closed-loop state instead of that in open-loop state is proposed to improve the operation rate of auto-locking, which only includes searching for transmission signal and optimizing the parameters. Using this method, it can avoid the influence of DC bias on the position of transmission peak signal in the process of switching from open-loop to closed-loop, which is beneficial to reduce the difficulty of searching for the transmission signal, and improve the probability of the successfully locking. This home-made controller is used in an ultra-stable laser stabilized to a 100-mm-long cavity, and the result shows that the median re-locking time of this ultra-stable laser is 16.6 s. By comparing with another ultra-stable laser, the median of the linewidth distribution was found to be 1.08 Hz, and the fractional frequency instability is less than less than $3.3\times10^{-15}$ within 100 s averaging time.

A schematic of the auto-locking ultra-stable laser is shown in figure 1(a) and the controller program is shown in the red dotted box. A commercial fiber laser (NKT Photonics Koheras Adjustik-E15) operating at 1550.12 nm was stabilized to an optical reference cavity by PDH technique. The laser frequency is locked to the Hermite-Gaussian 01 ($HG_{01}$) mode instead of the $HG_{00}$ mode due to nonnegligible losses in cavity, the finesse and coupling efficiency are 414000 and 11.2%, respectively. The cavity is made of the ultra-low expansion material (ULE) and placed horizontally on a U-shape aluminum block in a chamber with a vacuum of less $1\times10^{-6}$ Pa. By calculating with finite element, the cavity has a vibration sensitivity of less $1\times10^{-11}$/g. The vacuum chamber was temperature-stabilized at about 305 K with a fluctuation of below 10 mK during 60 hours. The vacuum and the all optical components for the PDH technique are placed on a commercial active vibration isolation platform with isolation degree of 20 dB.

This controller is developed with a combination of digital circuit and analog circuit. The digital circuit based on the FPGA is mainly used for diagnosing the operating state of the ultra-stable laser by output value of the PD2, and then controlling the action of servo devices liking offset of the PI1 and output value of voltage driver. The analog circuit is used for demodulating the discriminate signal and servo control, as to reduce the servo response time of the system. The acousto-optic (AOM) is used as the executor of the fast frequency servo and driven by a voltage controlled oscillator (VCO). The piezoelectric transducer (PZT) of this fiber laser is used as the executor of the slow frequency servo and driven by the voltage driver. The electro-optic modulator (EOM) is driven by a direct digital synthesizer (DDS), whose output frequencies are same and the phase of each channel can be adjusted from 0 to $2\pi$. The beat signal between the laser carrier and modulation sidebands is detected by the photodetector1 (PD1) and sent to the demodulator. The demodulated discriminate signal is sent to PI1 to control the output frequency of VCO and PI2 to control the output value of voltage driver. Here, the PI1 is developed with analog circuit, and its parameters will not change after being adjusted. The PI2 is based on the analog circuit and digital circuit, the analog circuit is used for servo control, and the digital circuit is used for adjusting the input signal offset until the output value of the PI2 is near zero. The 70% of the PD2 output value for HG01 mode is defined as the threshold in the auto-locking process to avoid the laser being locked to other modes.

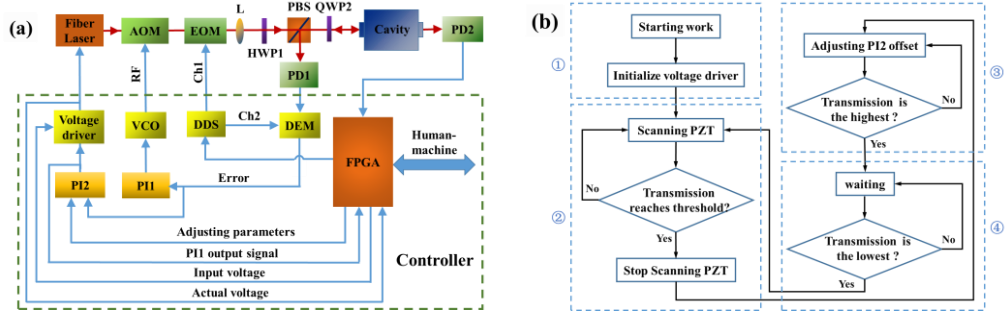

Fig.1. (a) Schematic of the auto-locking ultra-stable laser. AOM, acousto-optic modulator; EOM, electro-optic modulator; L, lens; HWP, half -wave plate; PBS, polarizing beam splitter; QWP, quarter-wave plate；PD1, PD2, Photodetector; VCO, voltage controlled oscillator; DDS, direct digital synthesizer; DEM, Demodulator; FPGA, Field programmable gate array; PI1, PI2, Proportional integral circuit. (b) Flow chart of automatic locking and relocking for ultra-stable laser

Figure 1(b) is the flow chart of the auto-locking. In the step 1, when the ultra-stable laser starts to operate, to find the transmission peak quickly the voltage driver outputs the initial set central output value 75V, which corresponds to the half free spectral range. In the step 2, the output value of the driver voltage changes as a triangular wave to scan laser PZT, the controller simultaneously diagnoses the output value of PD2 and executes corresponding commands. When the output value of PD2 reaches defined threshold range, the output value of the voltage driver stops changing and this value is recorded and set as the central voltage of the voltage driver. Then the frequency of laser is near to the resonance point of the reference cavity, the Servo1 starts to operate normally. In this step, 75V is taken as the central voltage value, and the scanning range is 10V, 20V, 30V, 40V… in turn. In the step 3, the controller adjusts the offset voltage of PI2 until the output value of the PD2 reaches the highest. In the step 4, when the ultra-stable laser operates normally, output value of the PD2 will be constantly monitored. The time interval of measured date is set as 1s to avoid misdiagnose caused by environment disturbance. If the output value of PD2 suddenly becomes the lowest suddenly, the controller will back to step 2. The output value of the voltage driver before losing lock will be set as the center voltage, and the lock point is near this output value. With the scanning method in step 2, the locking position of the transmission peak signal can be quickly found. Note that if the transmission signal is found to fall to the lowest in the step 3, the controller will also return to step 2.

When the ultra-stable laser stars to work, the temperatures of the fiber laser change from 298 K to 311 K. The discriminate signal and the transmission signal are recorded during auto-locking procedure shown by in figure 2(a) with blue line and red line, respectively, and the locking time of 153 seconds were taken. To ensure the repeatability of auto-locking after starting up, we simulate the auto-locking process from starting up to normally operation by changing the operating temperature of the fiber laser from the 311K to 298K, then to 311K without interruption, the locking time in these two processes are 144 s and 157 s, respectively. The results indicate that the ultra-stable laser after starting up can be stabilized within 160 s. The detail re-locking procedure after losing locking is shown in figure 2 (b), the red line and blue line represent the change of the error signal and the transmission signal in the process of relocking, respectively. The results show that the processes of searching for transmission signal and optimizing parameters take about 10 s and 6 s, respectively. To verify the repeatability of the controller, the ultra-stable laser is re-locked 157 times by periodically switching off the laser power via AOM, and the result is shown in figure (c). In bottom half of the figure 2(c), the histogram reveals that the probability for re-locking time from 16 s to 17 s is approximately 50%, and the computed median of the re-locking time distribution is about 16.6s. The top half of the figure 2 (c) shows that the probability for re-locking time less than 20s is 86% by integrating the histogram of re-locking time.

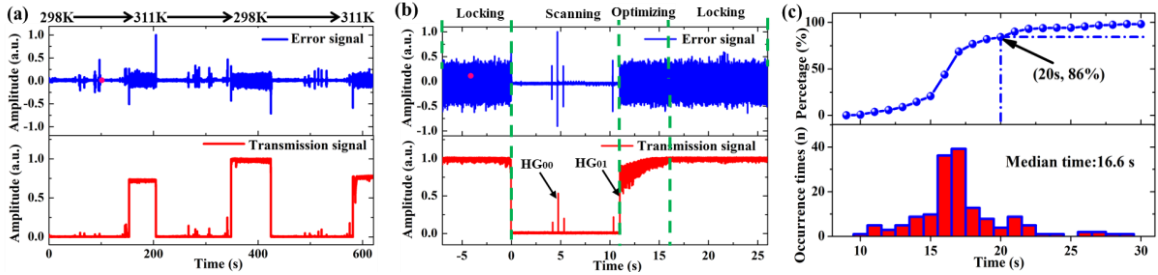

Fig. 2. (a) Auto-locking procedure after starting up. The blue line and red line represent the error signal and transmission signal, respectively. The 278 K is laboratory temperature and the 311 K is operating temperature of the fiber laser. (b) Auto-locking procedure after losing locking. (c) Distribution of the re-locking time for 157 groups. The blue line represents integrating the re-locking time, and the probability of re-locking time less than 20 s is 86%.

To ensure the performance of the auto-locking ultra-stable laser, we directly compare the beat note between this auto-locking ultra-stable laser and another one in our laboratory. The beat note is down-converted to 50 kHz from 726 MHz and recorded using a fast Fourier transformer (FFT) with a resolution bandwidth (RBW) of 0.5 Hz and a measurement time of 2 s. Each group of spectra was fitted with the Lorentzian function to obtain the linewidth of the beat note. To verify the reliable of the ultra-stable laser after re-locking, the laser was re-locked 100 times, and 10 groups of the spectra were measured each time. The distribution of the results is shown in figure 3(a), it reveals that the linewidth from 0.8 Hz to 1.2 Hz has a probability of about 63.4% showing at the bottom half and the computed median linewidth is about 1.08 Hz. The inset figure shows a typical full-width at half-maximum (FWHM) of 0.72 Hz (red line), which is the narrowest spectrum of a beat note observed. The top half shows the probability for linewidth of less than 1.5 Hz is 90% by integrating the linewidth. The beat note is directly measured by a frequency counter working on Λ-mode (Agilent 53230a). After removing a linear drift of about 0.42 Hz/s, the frequency instability of the stabilized laser is shown by black line in figure 3(b), it indicates that the fractional instability is approximately $1.8 \times 10^{-15}$ at the 1 s averaging time for ultra-stable laser, and rises to $3.4 \times 10^{-15}$ at 100 s averaging time. For short averaging times, the vibration and acoustic noise have a significantly effect on the instability. For long averaging times, the instability is mainly subjected to the locking point drift caused by temperature fluctuation and lower efficiency coupling leading a more sensitive locking point. The contribution of RAM to the instability is around $\sigma_1=5.6 \times 10^{-16}$ (blue line) at averaging time from 0.4 s to 20 s, the thermal noise floor of this cavity induced instability equals to $\sigma_2=6.3 \times 10^{-15}$ (pink dot line). The sum of these two additional noise sources is around $(\sigma_1^2+\sigma_2^2)^{1/2}=8.3 \times 10^{-16}$ at averaging time from 0.4 s to 20 s shown by red line.

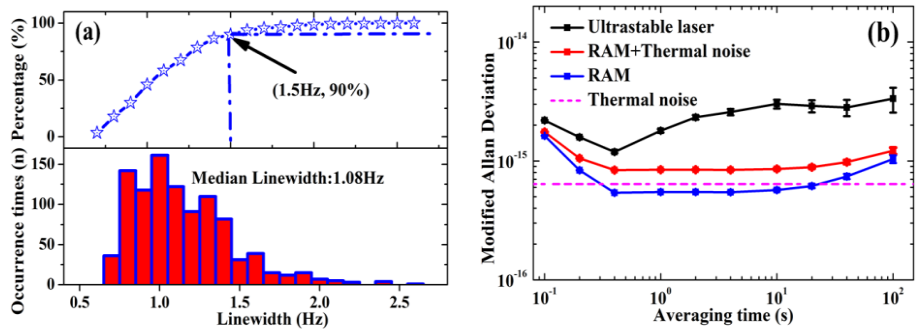

Fig.3. (a) Distribution of 1000 groups of laser line-width. The blue line represents integrating the line-widths, and the probability for linewidth of less than 1.5 Hz is 90%.The median of the linewidth distribution is 1.08 Hz, which is calculated by 1000 groups of the spectra. (b) Fractional frequency instability of the ultra-stable laser shown by black line. The blue line and pink line represent the RAM and thermal noise respectively. The red line is the effect of thermal noise and residual amplitude for ultra-stable laser.

In conclusion, an auto-locking ultra-stable laser is developed with a home-made controller, which is based on a combination of digital circuit and analog circuit. To reduce the locking time and improve the operation rate of auto-locking, an approach of searching for transmission signal in closed-loop state instead of that in open-loop state is proposed. And then, the performance and reliability of this auto-locking ultra-stable laser are tested. From 157 measurements without interruption, the computed median locking time is 16.6 s and the probability of locking time less than 20 s is 86%. By comparing with another ultra-stable laser, the linewidth distribution of 1000 groups of the spectra shows that the median linewidth is 1.08 Hz, and the probability of linewidth less than 1.5 Hz is 90%. The frequency instability is about $1.8 \times 10^{-15}$ at 1s averaging time, and rises to $3.4 \times 10^{-15}$ at 100 s averaging time. Next we will focus on the research of transportable ultra-stable laser to meet the application needs of users outside the laboratory.

## Acknowledgments

This work is supported by National Natural Science Foundation of China (NSFC) (11803041, 61127901, 91636101, 12103059 )

## References

[1] C. Q. Ma, L. F. Wu, Y. Y. Jiang, H. F. Yu, Z. Y. Bi, and L. S. Ma, Appl. Phys. Lett. 107, 261109 (2015).

[2] X. Deng, J. Liu, Q. Zang, D. D. Jiao, J. Gao, X. Zhang, D. Wang, R. Dong, and T. Liu, Appl. Opt. 32, 9620-9623 (2018).

[3] S. Droste, F. Ozimek, Th. Udem, K. Predehl, T. W. Hänsch, H. Schnatz, G. Grosche, and R. Holzwarth, Phys. Rev. Lett. 111, 110801 (2013).

[4] X. Deng, J. Liu, D. D. Jiao, J. Gao, Q. Zang, G. J. Xu, R. F. Dong, T. Liu, and S. G. Zhang, Chin. Phys. Lett. 33, 47-49 (2016).

[5] N. Chiodo, N. Quintin, F. Stefani, F. Wiotte, E. Camisard, C. Chardonnet, G. Santarelli, A. Amy-Klein, P. Pottie, and O. Lopez, Opt. Express 23, 33927 (2015)

[6] P. Kwee, C. Bogan, K. Danzmann, M. Frede, H. Kim, P. King, J. Pold, O. Puncken, R. L. Savage, F. Seifert, P. Wessels, L. Winkelmann, and B. Willke, Opt. Express 20, 10617 (2012).

[7] B. Willke, Laser Photonics Rev. 4, 780 (2010).

[8] S. Herrmann, A. Senger, K. Möhle, M. Nagel, E. Kovalchuk, and A. Peters, Phys. Rev. D 80, 105011 (2010).

[9] W. H. Oskay, W. M. Itano, and J. C. Bergquist, Phys. rev. lett. 94, 163001 (2005).

[10] M. D. Swallows, M. Bishof, Y. Lin, S. Blatt, M. J. Martin, A. M. Rey, and J. Ye, Science 331, 1043 (2011).

[11] N. Huntemann, M. Okhapkin, B. Lipphardt, S. Weyers, C. Tamm, and E. Peik, Phys. Rev. Lett. 108, 090801 (2012).

[12] S. Häfner, S. Falke, C. Grebing, S. Vogt, T. Legero, M. Merimaa, C. Lisdat, and U. Sterr, Opt. Lett. 40, 2112 (2015).

[13] W. Zhang, J. M. Robinson, L. Sonderhouse, E. Oelker, C. Benko, J. L. Hall, T. Legero, D. G. Matei, F. Riehle, U. Sterr, and J. Ye, Phys. Rev. Lett. 119, 243601 (2017).

[14] D. G. Matei, T. Legero, S. Häfner, C. Grebing, R. Weyrich, W. Zhang, L. Sonderhouse, J. M. Robinson, J. Ye, F. Riehle, and U. Sterr, Phys. Rev. Lett. 118, 263202 (2017).

[15] X. Y. Zeng, Y. X. Ye, X. H. Shi, Z. Y. Wang, K. Deng, J. Zhang, and Z. H. Lu, Opt. Lett. 43, 1690 (2018).

[16] T. Kessler, C. Hagemann, C. Grebing, T. Legero, U. Sterr, F. Riehle, M. J. Martin, L. Chen, and J. Ye, Nat. Photon. 6, 687 (2012).

[17] D. Kleppner, Phys. Today 59, 10–11, March (2006).

[18] C. W. Chou, D. B. Hume, T. Rosenband, and D. J. Wineland, Science 329, 1630 (2010).

[19] R. Schmidt, F. Flechtner, U. Meyer, K. Neumayer, C. Dahle, R. König, and J. Kusche, Surv. Geophys. 29, 319 (2008).

[20] W. Zhang, Z. Xu, M. Lours, R. Boudot, Y. Kersalé, G. Santarelli, and Y. L. Coq, Appl. Phys. Lett. 96, 211105 (2010)

[21] S. K. Lee , B. W. Han , and D. Cho. Rev. Sci. Instrum. 76, 026101 (2005).

[22] Y. Luo , H. Li , and H. C. Yeh, Rev. Sci. Instrum. 87, 056105(2016).

[23] http://www.stablelasers.com/products/integrated-1hz-stabilized-laser-systems/

[24] https://www.menlosystems.com/products/ultrastable-laser/sors-cubic/